\documentclass[]{aa}
\usepackage{graphicx}
\usepackage{txfonts}
\begin{document}
   \title{\ion{H}{i} Observations of an Ultra-Compact High-Velocity Cloud}

   \subtitle{}

   \author{C. Br\"uns 
   		\and
	   	T. Westmeier 
          }

   \offprints{C.Br\"uns,\\
    \email{cbruens@astro.uni-bonn.de}}

   \institute{Radioastronomisches Institut, Universit\"at Bonn,
   		Auf dem H\"ugel 71, D-53121 Bonn, Germany
             }

   \date{Received July 29, 2004; accepted September 6, 2004}

   \abstract{
   We present \ion{H}{i} observations of the compact high-velocity cloud \object{HVC289+33+251} that
   was discovered by Putman et al. (\cite{putman}). Observations with the 100-m Effelsberg telescope
   demonstrate that this cloud is still unresolved by the 9\arcmin\ beam of the Effelsberg
   telescope. The cloud shows a small line width of $\Delta v_{\rm FWHM} = 4.9$~km\,s$^{-1}$
   providing an upper limit to the kinetic temperature of the \ion{H}{i} gas of $T_{\rm k} \le 532$\,K.
   The total observed flux indicates an \ion{H}{i} mass of $M$(\ion{H}{i}) = 5.66$\cdot10^{4}$\,M$_\odot$~$\left[d/150\,{\rm kpc}\right]^2$.
   Follow-up \ion{H}{i} observations using the Australia Telescope Compact Array (ATCA) resolve 
   \object{HVC289+33+251} into 5 condensations that are embedded in a common \ion{H}{i} envelope. 
   The HVC shows a faint tail, indicating an ongoing ram-pressure interaction with an ambient low-density 
   medium.
   A FWHM diameter of $\vartheta$ = 4\farcm4 makes this HVC the by far most compact HVC known till now. 
   The observed parameters suggest that pressure stabilization by an ambient medium is rather unlikely.  
   At a distance of 150\,kpc, the virial mass is by a factor of 5.6 higher than the observed gas mass -- consistent 
   with \object{HVC289+33+251} being one of the ``missing'' dark matter mini halos that were predicted by 
   cosmological $\Lambda$CDM simulations (e.g. Klypin et al. \cite{klypin}; Moore et al. \cite{moore}). 
   Comparable clouds in other groups of galaxies or even around the Milky Way are not detectable with the 
   resolution and sensitivity of present surveys. 
   }

   \maketitle
%

\section{Introduction}

High-velocity clouds (HVCs) were first discovered by Muller et al. (\cite{muller}). They are defined as 
neutral atomic hydrogen clouds with radial velocities (relative to the local-standard-of-rest frame, LSR) that
cannot be explained by simple Galactic rotation models (Wakker \cite{wakkerdev}).   

Oort (\cite{oort}) proposed an extragalactic origin for HVCs. He argued, that the formation of galaxies 
is still an ongoing process and HVCs represent primordial clouds that are currently accreted by the
Milky Way. Blitz et al. (\cite{blitz}) revived the hypothesis that some HVCs are primordial gas left over 
from the formation of the Local Group galaxies. Braun \& Burton (\cite{bb99}) used the Leiden/Dwingeloo 
Survey (LDS) of neutral hydrogen (Hartmann \& Burton \cite{hartmann}) to compile a catalog of 66 
clouds with an angular diameter below 2\degr\ that are isolated from neighboring emission. They argued that 
this ensemble of compact HVCs (CHVCs) represents a homogeneous subsample of HVCs at Mpc distances. 
An HVC catalog of the southern sky with 1997 high-velocity objects and 179 CHVCs was compiled by 
Putman et al. (\cite{putman}) using the \ion{H}{i} Parkes All-Sky Survey (HIPASS, Barnes et al. \cite{barnes}).

Recent surveys (e.g. Zwaan \cite{zwaan}, de Blok et al. \cite{deblok}, Minchin et al. \cite{minchin}, 
Pisano et al. \cite{pisano}) failed to detect a similar population in other groups of galaxies. They concluded 
that CHVCs represent rather a circum-galactic population at typical distances below $d \le$ 150\,kpc 
than an intergalactic population at $d \approx$ 1\,Mpc. 

The CHVCs might represent the missing satellites expected from cosmological $\Lambda$CDM simulations 
(e.g. Klypin et al. \cite{klypin}; Moore et al. \cite{moore}). Recent models conclude that the majority of the 
dark matter mini halos is most likely located close to the Milky Way with a median distance of about 120\,kpc 
(e.g. Kravtsov et al. \cite{kravtsov}). Sternberg et al. (\cite{sternberg}) modeled the physical state of the 
\ion{H}{i} gas in dark matter mini halos and concluded that the observed CHVC parameters are consistent with a 
circum-galactic population at typical distances of $d \approx$ 150\,kpc. 

Several CHVCs have been studied in great detail (Braun \& Burton \cite{bb2000}, Br\"uns et al. \cite{bruens2}, 
Burton et al. \cite{burton}, de Heij et al. \cite{deheija}). These clouds have angular diameters larger than 30\arcmin\ 
and they show a complex morphology that could be explained by ram-pressure interaction with an ambient medium or
tidal interaction with the Milky Way. A CHVC that is undisturbed and spherically symmetric has not yet been found.

The compact cloud \object{HVC289+33+251} was discovered by Putman et al. (\cite{putman}, catalog number 1345).
\object{HVC289+33+251} is a CHVC that is unresolved by the effective Parkes beam of $\approx$15\farcm5. 
This CHVC is located $\approx$5\degr\ north of the Leading Arm of the Magellanic Clouds (Putman et al. \cite{putman98})
and shows a $\Delta v_{\rm LSR} \approx$ 50~km\,s$^{-1}$ higher radial velocity than the gas at the northern end of
the Leading Arm. No high-velocity \ion{H}{i} emission has been detected within a radius of 5\degr\ around \object{HVC289+33+251}, 
making it sufficiently isolated to be treated as a CHVC. 

In this paper, we present high-resolution \ion{H}{i} observations of \object{HVC289+33+251} using the 100-m Effelsberg
telescope and the Australia Telescope Compact Array (ATCA). In Sect.~2, we summarize these \ion{H}{i} observations
and outline the data reduction. In Sect.~3, we present the results of the single dish Effelsberg data and in Sect.~4 
those of the interferometric ATCA data. \object{HVC289+33+251} is considerably smaller than all other CHVCs that
have been studied so far. The extreme compactness compared to compact HVCs justifies the designation as an
ultra-compact HVC. The physical parameters of this potential dark matter mini halo are discussed in Sect.~5.


\section{Observations and Data Reduction}

\subsection{Effelsberg Data}

The observations were carried out in October 2002 with the \mbox{100-m} Effelsberg telescope using the 
21-cm receiver. The HPBW at 21-cm wavelength is 9\arcmin. The standard calibration source S7 was used for the 
flux calibration. 
The 1024 channel autocorrelator was split into two banks of 512 channels for the two orthogonal circular polarizations.
The bandwidth of 1.5 MHz offers a velocity resolution of $\Delta v$ = 0.65 $\rm km~s^{-1}$.
The HVC was mapped on a regular grid of 5 by 5 positions in equatorial coordinates with grid spacings of 
4\farcm5 and an integration time of 10 minutes for the central position and 3 minutes for the outer 
positions. This results in an rms-noise of $\sigma_{\rm rms}$ = 0.1~K and $\sigma_{\rm rms}$ = 0.17~K, respectively, 
after averaging the two polarizations.  We subtracted a third order polynomial for the baseline correction. 

\subsection{ATCA Data}

The high resolution \ion{H}{i} data were observed in February 2003 using the 750D configuration of
the ATCA interferometer. The five antennae provide baselines between 30\,m and 720\,m.
The correlator configuration FULL\_4\_1024-128 offers 1024 channels over a 
4~MHz bandwidth for both orthogonal polarizations. The velocity resolution is $\Delta v$ = 0.825 $\rm km~s^{-1}$. 
We used the source B1934-638 as primary calibrator and B1127-145 as secondary calibrator for the bandpass, 
gain and phase calibration. The CHVC was observed for 12 hours to allow for a 
good $uv$-coverage. The secondary calibrator was observed once per hour for 5 minutes.
The relatively northerly position of the source, combined with shadowing effects at low elevations
produced an elliptical synthesized beam of HPBW 112\farcs4$\times$35\farcs9 with a position angle of 2\fdg5. 

The data were reduced using the MIRIAD package (Sault et al. \cite{sault}). We have chosen robust weighting with a 
robustness parameter of 0.5 to optimize resolution and sensitivity. The deconvolution was performed using the 
maximum entropy method. The final data-cube has an rms-noise of \mbox{$\sigma_{\rm rms}$ = 7 mJy/beam}. 
We converted the observed fluxes, $S$, to brightness temperature, $T_{\rm mb}$, using
\begin{equation}
T_{\rm mb} = S \frac{\lambda^2}{2.65~\theta_{\rm maj}~\theta_{\rm min}},
\end{equation}
where $T_{\rm mb}$ and $S$ are measured in K and Jy\,beam$^{-1}$, respectively. $\lambda$ is the wavelength 
measured in cm, and $\theta_{\rm maj}$ and $\theta_{\rm min}$ are the sizes of the major and the minor beam 
in arc minutes. Using $\lambda$ = 21 cm and the beam size stated above yields a conversion factor of 
148.9 K(Jy\,beam$^{-1}$)$^{-1}$. The rms noise of the data-cube corresponds therefore to $\sigma_{\rm rms} \approx$ 1\,K 
using the brightness temperature scale.


\section{Results from the Effelsberg Telescope}\label{effdata}

\begin{figure}[t!]
\includegraphics[width=8.5cm]{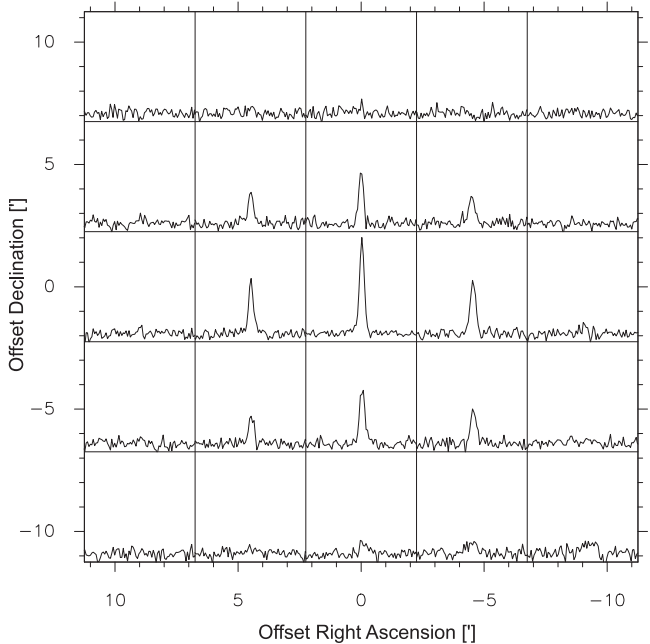}
\caption{Effelsberg \ion{H}{i} spectra of \object{HVC289+33+251}. The 4\farcm5 grid corresponds to HPBW/2 of
 the Effelsberg telescope. The spectra cover the velocity interval 200 km\,s$^{-1} \le v_{\rm LSR} \le$ 300 km\,s$^{-1}$ 
 and the brightness temperature interval --0.25 K $\le T_{\rm B} \le$ 3.2 K. Table~\ref{cores} lists the result of a 
 Gaussian fit to the central spectrum.}
\label{specmap}
\end{figure}

Figure~\ref{specmap} shows the 5 by 5 spectra map observed with the Effelsberg telescope. 
The 4\farcm5 grid corresponds to HPBW/2 of the Effelsberg telescope. The map demonstrates that 
\object{HVC289+33+251} is smaller than the beam size of 9\arcmin. There is, however, some faint 
emission detected in the south-west between offsets --9\arcmin\ $\le \Delta \alpha \le$ 0\arcmin\ at
$\Delta \delta$ = --9\arcmin. 
This emission has a column density of $N$(\ion{H}{i}) = (8$\pm$1)$\cdot$10$^{18}$\,cm$^{-2}$ and a FWHM line width
of $\Delta v_{\rm FWHM}$ = (13$\pm$2) km\,s$^{-1}$. It indicates a faint \ion{H}{i} tail associated with
\object{HVC289+33+251} (see Sect.~\ref{discussion}).
 
The central \ion{H}{i} spectrum has a peak intensity of $T_{\rm B,max}$~= (3.2$\pm$0.1)\,K, a line width of 
$\Delta v_{\rm FWHM}$~= (4.9$\pm$0.1)~km\,s$^{-1}$ and a column density of 
$N$(\ion{H}{i}) = (2.87$\pm$0.06)$\cdot$10$^{19}$\,cm$^{-2}$.
For comparison, Putman et al. (\cite{putman}) gave a peak intensity of $T_{\rm B,max}$~=~0.22\,K for
\object{HVC289+33+251}.
The considerable difference between $T_{\rm B,max,Eff}$ = 3.2 K and $T_{\rm B,max,PKS}$ = 0.22 K
is a result of the larger effective beam size of 15\farcm5 and the velocity resolution of the HIPASS data 
of 26.4 km\,s$^{-1}$ which is too coarse for a line width of \mbox{$\Delta v_{\rm FWHM} \approx$ 5 km\,s$^{-1}$}.
 
The total \ion{H}{i} mass of \object{HVC289+33+251} is related to the total \ion{H}{i} flux observed with the 
100-m Effelsberg telescope: 
\begin{equation}\label{gasmass}
\frac{M(\ion{H}{i})}{M_\odot} = 5301~\int S {\rm d}v~\left[\frac{d}{150\,{\rm kpc}}\right]^2,
\end{equation}
where $d$ is the distance to the source and $\int S {\rm d}v$ the integrated flux given in units of 
Jy\,km\,s$^{-1}$. The estimated mass depends on the unknown distance to the cloud. 
Using the observed total flux of $\int S {\rm d}v$ = (10.68$\pm$0.21)\,Jy\,km\,s$^{-1}$ yields an \ion{H}{i} mass of 
$M$(\ion{H}{i})~=~(5.66$\pm$0.11)$\cdot10^4$\,M$_\odot$~$\left[d/{\rm 150\,kpc}\right]^2$. 
The error reflects solely the observational uncertainty of the total flux.

\section{Results from the ATCA}\label{atcadata}

\begin{figure}[t!]
\centerline{
\includegraphics[width=8.9cm]{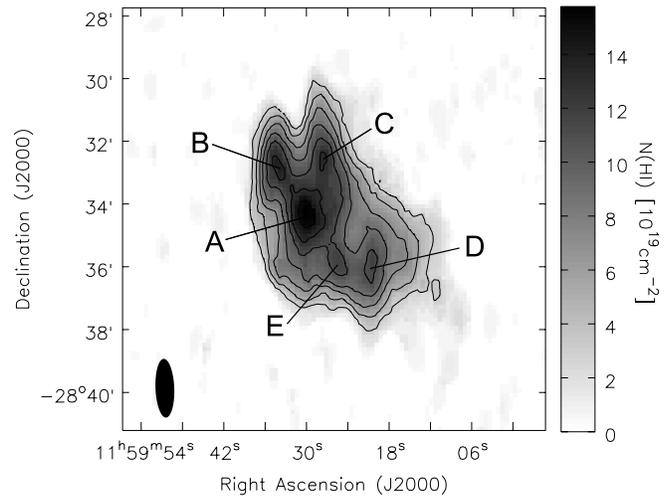}
}
\caption{\ion{H}{i} column density distribution of \object{HVC289+33+251} as observed with the ATCA. Contour lines
start at $N$(\ion{H}{i}) = 3$\cdot$10$^{19}$\,cm$^{-2}$ and increase in steps of 2$\cdot$10$^{19}$\,cm$^{-2}$.
The beam size is indicated in the lower left. The ATCA resolves the cloud into 5 concentrations A to E.}
\label{nhi}
\end{figure}

Figure~\ref{nhi} shows a column density map of the ATCA \ion{H}{i} data. The angular extent of the 
cloud is apparently smaller than the Effelsberg beam of HPBW = 9\arcmin. The angular resolution of 
the ATCA data of HPBW = 112\farcs4$\times$35\farcs9 is sufficiently high to resolve \object{HVC289+33+251}.
The angular diameter can be estimated by fitting a two-dimensional Gaussian to the column 
density distribution. The fit yields a FWHM diameter of $\vartheta$ = 4\farcm4. This is considerably
smaller than the typical sizes of compact HVCs with $\vartheta \ge$ 30\arcmin. The extreme compactness 
of \object{HVC289+33+251} compared to typical compact HVCs justifies the designation as an ultra-compact HVC.
The linear diameter, $D$, of \object{HVC289+33+251} is $D$~=~192\,pc~$\left[d/{\rm 150\,kpc}\right]$, where 
$d$ is the unknown distance to \object{HVC289+33+251}. 

The peak column density observed with the ATCA is $N$(\ion{H}{i})~=~1.9$\cdot$10$^{20}$\,cm$^{-2}$,
6.6 times higher than the column density observed with Effelsberg, clarifying the compactness of the HVC.
The total flux observed with the ATCA of $\int S {\rm d}v$ = 10.63\,Jy\,km\,s$^{-1}$ is in good
agreement with the observed single dish flux of $\int S {\rm d}v$ = 10.68\,Jy\,km\,s$^{-1}$.

The ATCA data resolve \object{HVC289+33+251} into 5 condensations that are embedded in a common envelope (Fig.~\ref{nhi}).
The coordinates of the five condensations are listed in Table~\ref{cores}. The central condensation (A)
is slightly more extended than the angular resolution. The angular diameter can be estimated by fitting a Gaussian
to the column density distribution in direction of right ascension, where the ATCA beam is HPBW = 35\farcs9.
The fit yields a diameter of FWHM = 96\arcsec\ which is consistent with an intrinsic diameter of about 
$\vartheta$ = 90\arcsec. The other 4 condensations are unresolved by the 112\farcs4$\times$35\farcs9 beam,
indicating that the intrinsic diameter must be smaller than $\vartheta$ $\le$ 50\arcsec. 
All five condensations have \ion{H}{i} column densities higher than $N$(\ion{H}{i})~$\ge$~1$\cdot$10$^{20}$\,cm$^{-2}$.

The spectra of the central positions of all five condensations were fit with Gaussians. The fits provide
the peak brightness temperature, the peak column density, the mean velocity, and the FWHM line width of the 
individual spectra. The results are summarized in Table~\ref{cores}. The five condensations have almost
the same mean velocity, i.e. there is no velocity gradient observed across the extent of \object{HVC289+33+251}.

A mean volume density of each condensation can be estimated assuming spherical symmetry for the condensations:
$\overline{n}_{\ion{H}{i}} = \frac{N(\ion{H}{i})}{\theta~d}$, where $\overline{n}_{\ion{H}{i}}$ is the mean density 
of a condensation, $N$(\ion{H}{i}) is the observed column density, $\theta$ is the angular diameter of the condensation 
given in radian, and $d$ is the distance to the cloud. The five condensations of \object{HVC289+33+251} have mean densities 
of the order $\overline{n}_{\ion{H}{i}}~\approx$~1\,cm$^{-3}$~$\left[d/{\rm 150\,kpc}\right]^{-1}$. 

The fit yields narrow line widths between $\Delta v_{\rm FWHM}$ = 3.8~km\,s$^{-1}$ for condensation B and
$\Delta v_{\rm FWHM}$ = 6.1~km\,s$^{-1}$ for condensation E (see Table~\ref{cores}). The line width provides an upper
limit to the kinetic temperature of the gas -- the so-called Doppler temperature, $T_{\rm D} = 21.8\cdot\Delta v_{\rm FWHM}^2$,
where $T_{\rm D}$ is measured in K and the observed line width, $\Delta v_{\rm FWHM}$, is measured in
units of km\,s$^{-1}$. This upper limit to the temperature is between $T_{\rm D}$ = 315\,K for condensation B and 
$T_{\rm D}$ = 811\,K for condensation E. Table~\ref{cores} also lists the values for the other condensations.

The upper limit for the temperature and the mean density can be used to derive an estimate of the
pressure in the condensations, using the approximation of an ideal gas, $P\,{\rm k}^{-1} = n\,T_{\rm D}$, 
where $P$ is the pressure of a condensation and k is the Boltzmann constant. The estimated pressures are between
$P\,{\rm k}^{-1}$ = 360 K\,cm$^{-3}$~$\left[d/{\rm 150\,kpc}\right]^{-1}$ for condensation A and 
$P\,{\rm k}^{-1}$ = 852 K\,cm$^{-3}$~$\left[d/{\rm 150\,kpc}\right]^{-1}$ for condensation E.
The gaseous condensations are expected to be located in a common gravitational potential filled with gas. 
This gas is expected to be in approximate pressure equilibrium. The median pressure of the condensations,
$P\,{\rm k}^{-1} \approx$ 450 K\,cm$^{-3}$~$\left[d/{\rm 150\,kpc}\right]^{-1}$, provides an estimate of the 
overall pressure of the gas in \object{HVC289+33+251}. 

Wolfire et al. (\cite{wolfire}) and Sternberg et al. (\cite{sternberg}) modeled the physical state of low-metalicity
HVCs. They concluded that pressures of the order few$\times$100~K\,cm$^{-3}$ are necessary to have a stable HVC 
consisting predominantly of a cold gas phase. Their pressures are consistent with our observationally determined value. 

\begin{table*}[t]
\caption{Parameters of the condensations in \object{HVC289+33+251} observed with the ATCA and the Effelsberg 
single-dish data. The columns give the name of the condensation, the telescope, the coordinates (see Fig.~\ref{nhi}),
the peak brightness temperature, $T_{\rm B,max}$, the peak column density, $N$(\ion{H}{i}), the mean velocity, 
$\overline{v}_{\rm LSR}$, the FWHM line width, $\Delta v_{\rm FWHM}$, the Doppler temperature, $T_{\rm D}$, the angular diameter, $\vartheta$,
the mean density, $\overline{n}_{\ion{H}{i}}$, and the estimated pressure, $P$\,k$^{-1}$. A distance of $d$ = 150\,kpc 
was assumed to derive the values for $\overline{n}_{\ion{H}{i}}$ and $P$\,k$^{-1}$.}
\begin{tabular}{lccccccccccc}
\hline
\hline
Region & Telescope & RA & Dec &  $T_{\rm B,max}$ & $N$(\ion{H}{i}) & $\overline{v}_{\rm LSR}$ & $\Delta v_{\rm FWHM}$ & $T_{\rm D}$ & $\vartheta$ & $\overline{n}_{\ion{H}{i}}$ & $P$\,k$^{-1}$\\
 & & J2000 & J2000 &  K & 10$^{20}$\,cm$^{-2}$ & km\,s$^{-1}$ & km\,s$^{-1}$ & K & \arcsec & cm$^{-3}$  & K\,cm$^{-3}$ \\
\hline
A & ATCA & 11$^{\rm h}$59$^{\rm m}$29\fs6 & --28\degr34\arcmin22\arcsec & 18.9 & 1.58 & 250.55 & 4.6 & 461 & 90 & 0.8 & 360\\
B & ATCA & 11$^{\rm h}$59$^{\rm m}$34\fs5 & --28\degr32\arcmin46\arcsec & 19.1 & 1.37 & 250.70 & 3.8 & 315 & $\le$50 & 1.2 & 384\\
C & ATCA & 11$^{\rm h}$59$^{\rm m}$27\fs8 & --28\degr32\arcmin38\arcsec & 15.6 & 1.33 & 250.03 & 4.7 & 482 & $\le$50 & 1.2 & 574\\
D & ATCA & 11$^{\rm h}$59$^{\rm m}$20\fs5 & --28\degr35\arcmin58\arcsec & 15.0 & 1.19 & 251.60 & 4.4 & 422 & $\le$50 & 1.1 & 447\\
E & ATCA & 11$^{\rm h}$59$^{\rm m}$25\fs4 & --28\degr35\arcmin50\arcsec & 10.9 & 1.18 & 251.65 & 6.1 & 811 & $\le$50 & 1.1 & 852\\
\hline
all & Effelsberg & 11$^{\rm h}$59$^{\rm m}$26\fs0 & --28\degr34\arcmin30\arcsec & 3.1 & 0.29 & 250.67 & 4.9 & 532 & 264 & 0.62& 330\\
\end{tabular}
\label{cores}
\end{table*}

\section{Discussion}\label{discussion}

Both data sets from the Effelsberg telescope (Fig.~\ref{specmap}) and the ATCA (Fig.~\ref{nhi}) indicate that 
there is a faint tail towards the south-west. A head-tail structure is a sign of ram-pressure interaction 
with an ambient medium (Br\"uns et al. \cite{bruens1},\cite{bruens2}; Quilis \& Moore \cite{quilis}; 
Konz et al. \cite{konz}). The numerical simulations from Quilis \& Moore (\cite{quilis}) and Konz et al. (\cite{konz})
suggest that the gas in an interacting HVC has its lowest temperatures at the leading edge, slightly higher 
temperatures in the head, and much higher temperatures in the tail of recently stripped material.
Exactly this trend is observed in terms of line widths (or Doppler temperatures): condensation B, that is located on the 
opposite side of the tail shows the lowest line width, $\Delta v_{\rm FWHM}$ = 3.8~km\,s$^{-1}$, the other condensations
have line widths of about $\Delta v_{\rm FWHM} \approx$ 4.6~km\,s$^{-1}$, while the gas in the tail has a much larger 
line width of $\Delta v_{\rm FWHM}$ = 13~km\,s$^{-1}$ (see Sect.~\ref{effdata}). These results indicate that 
\object{HVC289+33+251} is most likely embedded in a low-density medium, either in the outer Galactic halo or in the
intergalactic medium of the Local Group.

A distance of $d$ = 150\,kpc was assumed in the last two sections to derive the physical parameters: 
\ion{H}{i} mass, size, mean density, and pressure. The total \ion{H}{i} mass of \object{HVC289+33+251} is 
$M$(\ion{H}{i})~=~5.66$\cdot10^4$\,M$_\odot \left[d/150\,{\rm kpc}\right]^2$. The total gas mass must be higher as 
ionized hydrogen, helium, or molecular gas are not traced by the 21-cm line of neutral hydrogen:
\mbox{$M_{\rm gas}$ = $\mu \cdot M$(\ion{H}{i})}. The factor $\mu$ is the ratio of the total gas mass and the \ion{H}{i} mass. 
We expect to have 10 percent helium (by number) in this cloud, yielding $\mu \ge$ 1.4 and therefore 
\mbox{$M_{\rm gas} \ge$ 7.92$\cdot10^4$\,M$_\odot \left[d/150\,{\rm kpc}\right]^2$}.

\begin{figure}[t!]
\centerline{
\includegraphics[width=8.9cm]{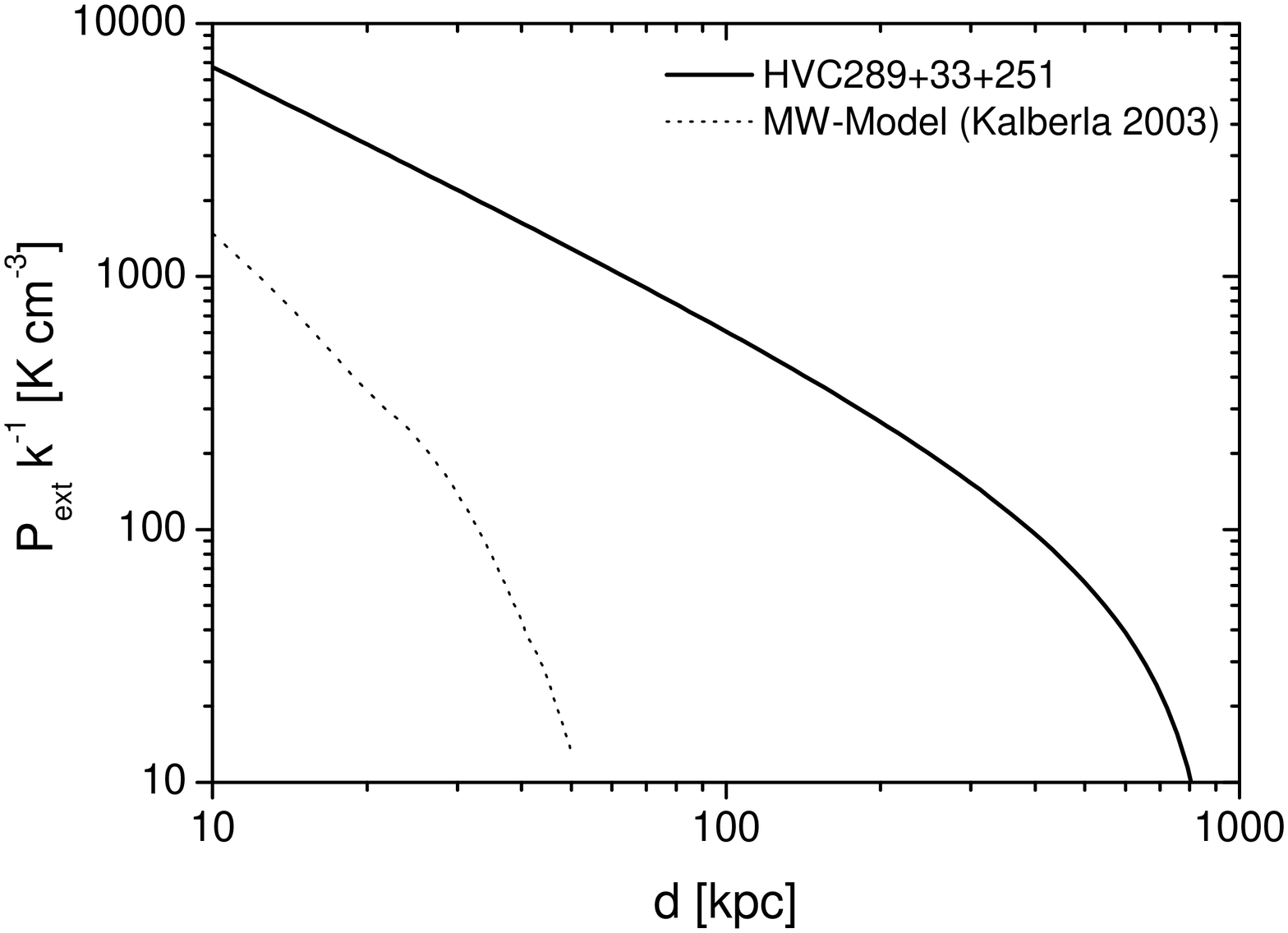}
}
\caption{The external pressure as a function of distance. The bold line represents the external pressure needed to 
stabilize \object{HVC289+33+251} according to Eq.~\ref{press}. The dotted line represents the pressure of the Milky Way halo
as given by the model of Kalberla (\cite{kalberla}). The pressure is calculated along the line-of-sight from the sun
to \object{HVC289+33+251}.}
\label{pressdist}
\end{figure}

The generalized version of the virial theorem for a homogeneous, spherically symmetric cloud including pressure support 
from an ambient medium is:
\begin{equation}
M_{\rm gas}~\overline{v^2} = \frac{3}{5} \frac{{\rm G} M_{\rm gas}^2}{R} + 3~P_{\rm ext}~V, \label{genvir}
\end{equation}
where $\overline{v^2}$ is the mean-square velocity that is related to the observed line-width by $\overline{v^2}$ = 3$\Delta v_{\rm FWHM}^2$/(8\,ln\,2).
G is the Gravitational constant, $R$ is the radius of the cloud given by \mbox{$R = \frac{\theta\,d}{2}$}, where $\theta$ is the 
angular diameter in radian and $d$ is the distance of the cloud. $P_{\rm ext}$ is the external pressure and 
$V = \frac{4}{3} \pi R^3$ is the volume of the cloud. 
Solving for the external pressure and using the overall line width from the Effelsberg data and the FWHM diameter of 
\object{HVC289+33+251} derived from the ATCA column density map (see Sect.~\ref{atcadata}) yields
\begin{eqnarray}
\frac{P_{\rm ext}}{{\rm k}} & = & 454~{\rm K~cm^{-3}} \left[\frac{d}{150\,{\rm kpc}}\right]^{-1} \left[\frac{\mu}{1.4}\right]
\left[\frac{\Delta v_{\rm FWHM}}{4.9~{\rm km\,s^{-1}}}\right]^2  \left[\frac{\theta}{4.4\arcmin}\right]^{-3} \nonumber \\
 & &  - 75~{\rm K~cm^{-3}} \left[\frac{\mu}{1.4}\right]^2 \left[\frac{\theta}{4.4\arcmin}\right]^{-4}. \label{press}
\end{eqnarray}
Figure~\ref{pressdist} shows the external pressure needed to stabilize \object{HVC289+33+251} as a function of distance to the sun, 
if its mass is given by its \ion{H}{i} mass including helium. For comparison, the dotted line indicates the
pressure in the halo of the Milky Way as given by the model of Kalberla (\cite{kalberla}).
This pressure is significantly lower than the pressure needed for stabilization at all distances.
Even an intergalactic medium with a very high pressure of $P\,k^{-1}$ = 200~K\,cm$^{-3}$ would not stabilize \object{HVC289+33+251} 
at distances less than $d \approx$ 250\,kpc (see Fig.~\ref{pressdist}).

The gas mass of \object{HVC289+33+251} could be higher, e.g. in form of a massive envelope of warm, ionized hydrogen. 
\object{HVC289+33+251} shows a faint head-tail structure, indicating an ongoing ram-pressure interaction with an ambient medium.
An outer envelope of warm, ionized hydrogen would be much more affected by this interaction. The fact that we see a head-tail
structure in the cold gas phase contradicts the existence of a massive envelope, as this envelope would have lost most of its mass
before the innermost regions were affected by the ram-pressure.

The assumption of virialization is not necessarily true as indicated by the faint \ion{H}{i} tail.
A dynamical time-scale can be estimated dividing the linear diameter of \object{HVC289+33+251} by its overall line-width.
Using the overall line width from the Effelsberg data and the FWHM diameter derived from the ATCA column density map 
yields a time-scale of $\tau$ = 38.3 Myr $\left[d/150\,{\rm kpc}\right]$. 
One orbit around the Milky Way is expected to take a few Gyrs, i.e. the gravitational forces from the Milky Way and 
ram-pressure forces should be approximately constant over considerably longer time-scales than 
$\tau$ = 38.3 Myr $\left[d/150\,{\rm kpc}\right]$. This time-scale is therefore sufficiently short to assume that 
\object{HVC289+33+251} is at least close to equilibrium.

The assumption of virialization with negligible external pressure allows us to estimate the mass needed to stabilize 
\object{HVC289+33+251} at a given distance. The virial mass for a $n~\sim~R^{-1}$ density distribution is
\begin{equation}
M_{\rm vir} = 190~R~\Delta\,v_{\rm FWHM}^2,
\end{equation}
where $M_{\rm vir}$ is the virial mass in units of solar masses, $\Delta v_{\rm FWHM}$ is the line width in units of 
km\,s$^{-1}$, and $R$ is the radius of the cloud in units of pc.
Using the same parameters as in Eq. \ref{press} yields a virial mass of 
\begin{equation}
M_{\rm vir}~=~4.4\cdot10^5\,M_\odot \left[\frac{d/150}{{\rm kpc}}\right] \left[\frac{\theta}{4.4\arcmin}\right] \left[\frac{\Delta v_{\rm FWHM}}{4.9~{\rm km\,s^{-1}}}\right]^2.
\end{equation}
The ratio of the virial mass and the gas mass is
\begin{equation}
\frac{M_{\rm vir}}{M_{\rm gas}} = 5.6 \left[\frac{d}{150\,{\rm kpc}}\right]^{-1} \left[\frac{\mu}{1.4}\right]^{-1}.
\end{equation}
The distance where the gas mass equals the dynamical mass, $d$~=~825\,kpc, can be used as an upper limit of the 
distance to \object{HVC289+33+251} (see also Fig. \ref{pressdist}). 
While no dark matter is needed for stabilization at $d$~=~825\,kpc, about 80 percent of the dynamical mass would be 
undetected at a distance of $d$ = 150\,kpc. For comparison, about 95 percent dark matter would be needed for 
stabilization at a distance of $d$ = 50\,kpc. These percentages of dark matter are common for various types of galaxies. 

Stabilization by dark matter yields a reasonable amount of dark matter, while the stabilization by the pressure of an 
ambient medium or a massive envelope of ionized hydrogen is very unlikely. The true distance to \object{HVC289+33+251} -- 
the key parameter for the physical state -- cannot be derived on the basis of the data presented in this paper.
The observed morphology and the estimated physical parameters of the compact cloud \object{HVC289+33+251} are, however, 
consistent with a low-mass dark matter mini halo in the vicinity of the Milky Way that moves through a low density medium.

The FWHM diameter of only 4\farcm4 makes this cloud undetectable for low-resolution telescopes like the 25-m Dwingeloo 
telescope. Its angular resolution of HPBW = 36\arcmin\ yields a beam filling factor of about 1.5\%.
The considerable difference between the peak brightness temperatures observed with Effelsberg 
($T_{\rm B,max,Eff}$ = 3.2\,K) and the HIPASS survey using the Parkes telescope ($T_{\rm B,max,PKS}$ = 0.22\,K)
clarifies the importance of high angular {\em and} velocity resolution for the detection of clouds like 
\object{HVC289+33+251}. An assumed distance of $d$~=~150\,kpc corresponds to a linear diameter of 192\,pc.
A comparable cloud located in a nearby group at $d \approx$ 3\,Mpc  would have an angular diameter of about $\vartheta$ 
$\approx$ 13\arcsec\ -- the size of the synthesized beam of the VLA in the C~configuration. A signal of that size 
that appears only in one spectral channel would most likely be regarded as a noise peak.

Recent surveys for compact clouds in galaxy groups, e.g. Zwaan (\cite{zwaan}), de Blok et al. (\cite{deblok}), 
Minchin et al. (\cite{minchin}), Pisano et al. (\cite{pisano}), have mass detection limits of 
$M$(\ion{H}{i}) = 7$\cdot$10$^6$M$_\odot$, $M$(\ion{H}{i}) = 3$\cdot$10$^6$M$_\odot$, 
$M$(\ion{H}{i}) = 2$\cdot$10$^6$M$_\odot$, and $M$(\ion{H}{i})~=~4$\cdot$10$^5$M$_\odot$, respectively.
They all lack the spatial resolution and sensitivity to detect clouds comparable to \object{HVC289+33+251} in these
galaxy groups. 

A larger sample of similar clouds in the vicinity of the Milky Way and other galaxies might 
exist that are undetectable for currently available surveys.

\begin{acknowledgements}
      T. Westmeier is supported by the German \emph{Deut\-sche For\-schungs\-ge\-mein\-schaft, DFG\/} project
      number KE 757/4-1. Based on observations with the 100-m telescope of the MPIfR (Max-Planck-Institut f\"ur 
      Radioastronomie) at Effelsberg. The ATCA is part of the Australia Telescope which is funded by the Commonwealth 
      of Australia for operation as a National Facility managed by CSIRO.
\end{acknowledgements}


\begin{thebibliography}{} 
 \bibitem[2001]{barnes} Barnes, D. G., Staveley-Smith, L., de Blok, W. J. G., et al. 2001, \mnras, 322, 486
 \bibitem[1999]{blitz} Blitz, L., Spergel, D.N., Teuben, P.J., Hartmann, D., Burton, W. B. 1999, \apj, 514, 818
 \bibitem[2002]{deblok}	de Blok, W. J. G., Zwaan, M. A., Dijkstra, M., Briggs, F. H., Freeman, K. C. 2002, \aap, 382, 43
 \bibitem[1999]{bb99} Braun, R., Burton, W.B. 1999, \aap, 341, 437
 \bibitem[2000]{bb2000}  Braun, R., Burton, W. B. 2000, \aap, 354, 853
 \bibitem[2000]{bruens1}  Br\"uns, C., Kerp, J., Kalberla, P.M.W., Mebold, U. 2000, \aap, 357, 120
 \bibitem[2001]{bruens2}  Br\"uns, C., Kerp, J., Pagels, A. 2001, \aap, 370, L30
 \bibitem[2001]{burton} Burton, W.B., Braun, R., Chengalur, J.N. 2001 \aap, 369, 616
 \bibitem[1997]{hartmann} Hartmann, D., Burton, W. B. 1997, ``An atlas of Galactic Neutral Hydrogen Emission'', Cambridge University Press
 \bibitem[2002a]{deheija} de Heij, V., Braun, R., Burton, W. B. 2002, \aap, 391, 67
 \bibitem[2003]{kalberla}  Kalberla, P.M.W. 2003, \apj, 588, 805
 \bibitem[1999]{klypin} Klypin, A., Kravtsov, A.V., Valenzuela, O., Prada, F. 1999, \apj, 522, 82
 \bibitem[2004]{kravtsov} Kravtsov, A.V., Gnedin, O.Y., Klypin, A.A. 2004, \apj, 609, 482
 \bibitem[2002]{konz} Konz, C., Br\"uns, C., Birk, T. 2002, \aap, 391, 713
 \bibitem[2003]{minchin} Minchin, R. F., Disney, M. J., Boyce, P. J., et al. 2003, \mnras, 346, 787
 \bibitem[1999]{moore} Moore, B., Ghigna, S., Governato, F., et al. 1999, \apj, 524, 19
 \bibitem[1963]{muller} Muller, C.A., Oort, J.H., Raimond, E. 1963, C.R. Acad. Sci. Paris, 257, 1661
 \bibitem[1966]{oort} Oort, J.H. 1966, Bull. Astr. Inst. Netherlands, 18, 421
 \bibitem[2004]{pisano}	Pisano, D. J., Barnes, D.G., Gibson, B.K., et al. 2004, \apj, 610, 17 
 \bibitem[1998]{putman98} Putman, M.E., Gibson B.K., Staveley-Smith L., et al. 1998, \nat, 394, 752
 \bibitem[2002]{putman} Putman, M.E., de Heij, V., Staveley-Smith, L., et al. 2002, \aj, 123, 873 
 \bibitem[2001]{quilis} Quilis, V., Moore, B. 2001, \apj, 555, L95
 \bibitem[1995]{sault} Sault, R. J., Teuben, P. J., Wright, M. C. H., 1995, in ASP Conf. Ser. 77, Astronomical Data 
 Analysis Systems IV, ed. R. A. Shaw, H. E. Payne, \& J. J. E. Haynes (San Francisco: ASP), 433
 \bibitem[2002]{sternberg} Sternberg, A., McKee, C.F., Wolfire, M.G. 2002, \apjs, 143, 419
 \bibitem[2001]{zwaan}	Zwaan, M.A. 2001, \mnras, 325, 1142
 \bibitem[1991]{wakkerdev} Wakker, B. P. 1991, \aap, 250, 499
 \bibitem[1995]{wolfire} Wolfire, W.G., McKee, C.F., Hollenbach, D., Tielens, A.G.G.M. 1995, \apj, 453, 673
\end{thebibliography}
\end{document}